\documentclass[twocolumn,preprintnumbers,amsmath,nofootinbib,amssymb]{revtex4}
\usepackage{amssymb}
\usepackage{latexsym}
\usepackage{color}
\usepackage{enumerate}
\usepackage{graphicx}
\usepackage{epsfig}
\usepackage{hyperref}
\usepackage{color}

\def\nn{\nonumber\\}
\newcommand{\f}[2]{\frac{#1}{#2}}
\def\be{\begin{equation}}
\def\ee{\end{equation}}
\def\bea{\begin{eqnarray}}
\def\eea{\end{eqnarray}}
\begin{document}
\title{Structure Formation in Generalized Rastall Gravity}
\author{$^1$A. H. Ziaie\footnote{ah.ziaie@maragheh.ac.ir}, $^1$H. Moradpour\footnote{h.moradpour@maragheh.ac.ir}, $^2$H. Shabani\footnote{h.shabani@phys.usb.ac.ir}}
\address{$^1$Research Institute for Astronomy and Astrophysics of Maragha (RIAAM), University of Maragheh, P.O. Box 55136-553, Maragheh, Iran\\}
\address{$^2$Physics Department, Faculty of Sciences, University of Sistan and Baluchestan, Zahedan, Iran}
\begin{abstract}
Recently a modified version of Rastall theory of gravity has been introduced in which a varying coupling parameter could act as dark energy (DE) and thus, it can be held responsible for the current accelerated expansion of the Universe. Motivated by this modification, we study here the evolution of linear and non-linear perturbations in the matter content of the Universe, utilizing spherically symmetric top-hat collapse scenario. The exact solutions we obtain in linear regime show that as the Universe evolves, matter density perturbations grow and reach a maximum value at a certain redshift after which these perturbations start decreasing towards a finite positive value at the present time. Depending on model parameters, exact oscillatory solutions can be also found representing that matter perturbations could experience either overdense and underdense regions during the dynamical evolution of the Universe. Numerical solutions in non-linear regime show that the amplitude of perturbations grow much faster than the linear one and diverges at a critical redshift. However, the formation of collapsed structures is delayed as compared to $\Lambda${\rm CDM} model. It is found that the running mutual interaction between matter and geometry, encoded in the variable Rastall coupling parameter, could drastically affect the dynamics of matter perturbations and their growth rate during the evolution of the Universe.
\end{abstract}

\maketitle
\section{Introduction}
About two decades since the discovery of cosmic acceleration found by observational data~\cite{cosacc,laterconf}, many efforts have been done in order to figure out the physical mechanism behind this phenomenon. In this sense, many cosmological models propose a new and mysterious component within the Universe, the so-called DE~\cite{demodels}. Though much attempts that have been made so far, the physical nature of DE is still a mystery to us. The most simple and famous candidate for this unknown energy is the cosmological constant $\Lambda$, a perfect fluid whose equation of state parameter obeys $w=p/\rho=-1$, which is a constant throughout the cosmic evolution~\cite{coscons}. Also, a dynamical approach to the problem of DE has been introduced in which the observed accelerated expansion of the Universe originates from the potential energy of a scalar field, referred to as quintessence field. For recent reviews we consult the reader to~\cite{dynDE}. Apart from observing DE as a real ingredient of our cosmos, one possible way to address the issue of cosmic speed-up is to modify gravitation theory such that the accelerated expanding phase could be attributed to this modification. Some remarkable examples in this regard are scalar-tensor theory~\cite{BDacccos}, Kinetic Gravity Braiding (KGB)~\cite{CosKGB} and $f(R)$ models~\cite{CosfR}, see also~\cite{grmodde} for recent reviews.
\par
The energy-momentum source in General theory of relativity (GR) and most of its modifications is described with the help of a divergence-free tensor which has a minimal coupling to the spacetime geometry~\cite{grmodde}. However, in~\cite{ppp1,ppp}, it is argued that this property of material source, which leads to the continuity of energy-momentum tensor (EMT), i.e., $\nabla_\nu T^{\mu\nu}=0$, does not hold for the process of particle production. Hence, It makes sense to put aside the condition on EMT conservation and look for a novel gravitation theory. In 1972, Peter Rastall proposed such a modified gravity theory wherein, the conservation of EMT in curved spacetime is questioned, as this law is examined only in the flat Minkowski spacetime or particularly in the limit of weak gravitational fields~\cite{rastall1972}. Based on Rastall's argument, the vanishing of covariant derivative of EMT does no longer hold and instead one assumes that this vector field is proportional to the gradient of the Ricci curvature scalar, i.e., $\nabla_\nu  T^{\mu\nu}\propto\nabla^\mu R$. Indeed, this theory could provide a phenomenological setting for discerning aspects of quantum effects within gravitational interactions, i.e the violation of conservation laws in the realm of classical physics~\cite{ppp1,RastallQuantum,fabris2015}. During the past years, many efforts have been devoted with the aim of exploring different aspects of Rastall gravity among which we can quote: matter-geometry non-minimal coupling~\cite{rastall1972,RastallQuantum}, compatibility with observational probes on the age of Universe and on the Hubble parameter~\cite{FESRASPLB}, helium nucleosynthesis~\cite{nuclras}, structure of neutron stars~\cite{NeutRast} and the ability of theory to explain the accelerated expansion phase and inflationary problems~\cite{Rassaccinf}, see also~\cite{Akarsu2020} for a comprehensive and detailed study of the observational aspects of Rastall gravity. Moreover, compared to GR, a better description for matter dominated epoch can be rendered in Rastall theory~\cite{Effemtras}. Also, motivated by these evidences, physicists have tried to investigate various cosmic epochs in this framework~\cite{rascosmiceras} and in~\cite{Rasentrpr}, it is argued that this theory does not suffer from the entropy and age problems that appear in the context of standard cosmology.
\par
An intriguing problem in cosmology is the beginning and evolution of cosmological structures such as galaxies and galaxy clusters observed in large surveys. The most commonly accepted answer to this question is provided by the gravitational instability scenario in which, galaxies, galaxy clusters, and galaxy super-clusters all arise from gravitational instability that amplifies very small initial density fluctuations during the evolution of Universe. Such fluctuations then grow slowly over time till they become robust enough to get detached from the background expansion and finally collapse into gravitationally bound systems like galaxies and galaxy clusters. Therefore, taking into account the initial seed for gravitational instabilities, the task is to present, at least, analytic forecasts to illustrate the gravitational assembling of a huge amount of matter that gives rise to the formation of large-scale structures we observe today~\cite{white1978,peebles1993,peacock 1999}. Apart from its unavoidable role in the current cosmic speed-up, the DE component could have its own effects on the dynamics and final fate of cosmological perturbations. Hence, searching for the footprints of DE through its influences on the formation and growth rate of cosmic structures is of significant importance. A convenient model to investigate the structure formation in the presence of DE is provided by Top-Hat Spherical-Collapse (SC) model~\cite{SCintrodu} which was initially employed for Einstein-de Sitter (EdS) background in the standard cold-dark-matter scenario~\cite{tophatscdm}, and later in $\Lambda$CDM~\cite{THLambdacdm}. Work along this line has also been extended to other cosmological settings such as, quintessence fields~\cite{TopHqfield}, decaying vacuum models~\cite{TopHdv}, $f(R)$ gravity theories~\cite{fRTopH}, cosmological models with constant equation of state for DE~\cite{abramo2007,DEconseqsTH} and coupled DE models~\cite{coupDETH}.
\par
Recently, a generalization of Rastall theory has been proposed in which the Rastall parameter is taken to be a variable during the dynamical evolution of the Universe~\cite{EPJCMorad}. The authors found that a running coupling between pressure-less matter and geometry can play the role of DE and hence the current accelerating cosmic phase may have been originated from such a dynamic matter-geometry interaction. In~\cite{das2018}, it is shown that the initial singularity of the Universe can be avoided considering a suitable functionality of Rastall parameter. Moreover, a complete cosmic scenario can be achieved from early inflationary phase to the current accelerating stage through the matter dominated era for this model of gravity~\cite{das2018}. Motivated by these considerations, our aim here is to consider the issue of structure formation in generalized Rastall gravity and investigate the possible consequences of a varying Rastall parameter on the growth of density perturbations. The paper is then organized as follows. In Sec.~\ref{fesras} we briefly review the field equations of generalized Rastall
gravity. Utilizing the (SC) model, in Sec.~\ref{maineqs}, we derive the evolutionary equations governing pressure-less matter perturbations. Sec.~\ref{sollineareqs} deals with exact solutions to the perturbation equation in linear regime. In Sec.~\ref{nonlinregime} we investigate the evolution of matter perturbations in non-linear regime and finally in Sec.~\ref{conclrems} we summarize our results.
\section{Field equations of Generalized Rastall gravity}\label{fesras}
Rastall assumed that the coupling coefficient between divergence of EMT and the derivative of Ricci scalar, called Rastall parameter, is constant~\cite{rastall1972}, a hypothesis meaning that the Universe evolution does not affect the rate of energy-momentum transfer between matter fields and geometry~\cite{EPJCMorad}. Moreover, this theory cannot remove the need to a mysterious DE source in order to describe the accelerated
Universe \cite{rascosmiceras1}. In this regard, a simple
generalization of Rastall theory as \be\label{CoVDiv}
\nabla_{\mu}T^{\mu}_{\,\,\,\nu}=\nabla_{\nu}(\lambda R), \ee in
which the Rastall parameter $\lambda$ is no longer a constant, can
describe the accelerated Universe without needing to a DE
source \cite{EPJCMorad}. The field equations are as follows
\be\label{genrasfieldeqs} G_{\mu\nu}+\kappa\lambda
g_{\mu\nu}R=\kappa T_{\mu\nu}, \ee where $\kappa$ is Rastall
gravitational constant \cite{EPJCMorad}.
\section{Spherical collapse}\label{maineqs}
The line element for a spatially flat, homogeneous and isotropic Universe is given by
\be\label{lineelement}
ds^2=-dt^2+a(t)^2\left[dr^2+r^2(d\theta^2+\sin^2\theta)d\phi^2\right],
\ee
where $a(t)$ is the scale factor of the Universe. The energy-momentum source filling the Universe is assumed to be a perfect fluid with an isotropic {\rm EMT} given by
\be\label{emtsource}
T_{\mu}^{\,\nu}={\rm diag}[-\rho(t),p(t),p(t),p(t)],
\ee
where $\rho(t)$ is the energy density and $p(t)$ is the isotropic pressure. The Friedmann equations can be put into the form
\bea
(12\kappa\lambda-3)H^2+6\kappa\lambda\dot{H}=-\kappa\rho,\label{fieldback}\\
(12\kappa\lambda-3)H^2+(6\kappa\lambda-2)\dot{H}=\kappa
p,\label{fieldback1}, \eea where $H=\dot{a}/a$ and an over-dot
denotes derivative with respect to cosmic time $t$. Considering a
dust fluid as the only constituent of the Universe, i.e.,
$p=p_m=0$, the conservation equation (\ref{CoVDiv}) leads to the
following continuity equation in generalized Rastall gravity, as
\be\label{conteqH}
g(\lambda)\dot{\rho}_m+\rho_m\left[3H-\dot{f}(\lambda)\right]=0,
\ee 
where 
\bea\label{fg}
g(\lambda)=\f{3\kappa\lambda-1}{4\kappa\lambda-1},~~~~f(\lambda)=\f{\kappa\lambda}{4\kappa\lambda-1}.
\eea We note that in GR limit where
\be\label{GRL}
\lambda\rightarrow0,~~~~g(\lambda)\rightarrow1,~~~~f(\lambda)\rightarrow0,
\ee
the usual continuity equation will be recovered. Let us now
consider a spherically symmetric region of radius $r$ filled with
a dust cloud of homogeneous density $\rho^{{\rm c}}_{m}$. In the
framework of {\rm SC} model, this region is described by a top-hat
profile and uniform density so that at time $t$, $\rho^{{\rm
c}}_{m}(t)=\rho_{m}(t)+\delta\rho_{m}$. In other words, this
region initially experiences a small perturbation of the
background fluid density, i.e., $\delta\rho_{m}$ and is immersed
within a homogeneous Universe with energy density $\rho_{m}$. If
$\delta\rho_{m}>0$ the spherical region will finally collapse
under its own gravitational weight, otherwise, it will expand
faster than the average Hubble flow, generating thus, what is
known as a void. In analogy with Eq.~(\ref{conteqH}), the
continuity equation for the spherical region can be written as
\be\label{conteqh} g(\lambda_{\rm c})\dot{\rho}_m^{\rm
c}+\rho_m^{\rm c}\left[3h-\dot{f}(\lambda_{\rm c})\right]=0, \ee
where, $h=\dot{r}/r$ denotes the local expansion rate inside the
spherical perturbed region. In general, the Rastall parameter can
be different within the spherical region and outside of it,
however, for the sake of simplicity we assume that this parameter
takes the same values for the local and background regions, i.e.,
$\lambda_{\rm c}=\lambda$. A useful quantity through which the
evolution of perturbations can be better investigate is the
density contrast of a the fluid which is defined as
\be\label{dencontd} \delta_m=\f{\rho^{\rm
c}_m}{\rho_m}-1=\f{\delta\rho_m}{\rho_m}. \ee Indeed this quantity
measure the deviation of the local fluid density from the
background density. In order to calculate the evolution of density
contrast, we take the time derivative of Eq.~(\ref{dencontd})
giving \bea\label{firstderdelta}
\dot{\delta}_m=\f{3}{g}(H-h)(1+\delta_m), \eea where use has been
made of conservation equations (\ref{conteqH}) and
(\ref{conteqh}). Differentiating again with respect to time gives
\bea\label{dencontdtdt}
\ddot{\delta}_m=-\f{\dot{g}}{g}\dot{\delta}_m+\f{3}{g}(1+\delta_m)(\dot{H}-\dot{h})+\f{\dot{\delta}^2_m}{1+\delta_m}.
\eea 
In order to estimate the second term
of the above equation we consider the second Friedmann equation
for the local and background regions, given as
\bea\label{secondfeqHh}
\f{\ddot{a}}{a}=-\f{\kappa(1-6\kappa\lambda)}{6(1-4\kappa\lambda)}\rho_m,~~~~\f{\ddot{r}}{r}=-\f{\kappa(1-6\kappa\lambda)}{6(1-4\kappa\lambda)}\rho_m^{\rm
c}, \eea whence we have \be\label{Hdot-hdot}
\dot{H}-\dot{h}=\f{\kappa(1-6\kappa\lambda)}{6(1-4\kappa\lambda)}\rho_m\delta_m+h^2-H^2.
\ee Substituting the above relation into equation
(\ref{dencontdtdt}) along with using the first derivative of
density contrast, Eq.~(\ref{firstderdelta}) we finally get
\bea\label{ddotdeltam}
\ddot{\delta}_m&+&\left(2H+\f{\dot{g}}{g}\right)\dot{\delta}_m-\f{(g+3)\dot{\delta}^2_m}{3(1+\delta_m)}\nn&-&\f{\kappa(1+\delta_m)(1-6\kappa\lambda)\rho_m\delta_m}{2g(1-4\kappa\lambda)}=0.
\eea We note that for $\lambda=0$, the above equation reduces to
its counterpart for a single dust fluid given
in~\cite{abramo2007}. In order to study the evolution of matter
perturbations we need to determine the functionality of the
Rastall parameter. To this aim, we assume that the matter energy density obeys that of a pressure-less matter which its dependence on scale factor is given by, $\rho_m(a)=\rho_{0m}a^{-3}$, where $\rho_{0m}=\rho_m(a=1)$. Substituting then for the energy density into the conservation
equation~(\ref{conteqH}), we obtain the behavior of $\lambda(a)$
parameter as
\be\label{lambdaa}
\lambda(a)=\f{1}{\kappa(4+\alpha a^{-3})},~~~~\alpha=C\rho_{0m},
\ee
where $C$ is a constant of integration. For further discussions and details we consult the reader to~\cite{EPJCMorad}. Indeed, for
$C=0$ (or even, $a\gg1$) we have $\lambda=1/4\kappa$ that
guarantees the existence of inflationary phases for the Universe
expansion~\cite{EPJCMorad}. We further note that using the above
solution along with Eq. (\ref{fieldback}) we get the behavior of
Hubble parameter as \be\label{Hubblepar}
H(a)=\f{\kappa\rho_{0m}}{3\alpha}\left[\f{\alpha+a^3}{a^3}\right]^{\f{1}{2}},
\ee whence we get the deceleration parameter as \be\label{decpara}
q=-1-\f{\dot{H}}{H^2}=-1+\f{1+z}{H(z)}\f{dH(z)}{dz}=\f{\alpha(1+z)^3-2}{2(1+\alpha(1+z)^3)},
\ee where $a=1/(1+z)$ and $z$ being the redshift. To obtain the
behavior of density contrast as a function of redshift we first
transform the time derivatives in Eq. (\ref{ddotdeltam}) to
derivatives with respect to the scale factor, bearing in mind the
following relations \bea\label{ttoa}
\ddot{\delta}_m=a^2H^2\delta^{\prime\prime}+aH^2\left[\f{2g-3}{2g}\right]\delta_m^\prime,~~~~\dot{\delta}_m=aH\delta_m^\prime,
\eea where a prime denotes derivative with respect to the scale
factor and use has been made of the second Friedmann equation
(\ref{fieldback1}). Substituting for the derivatives into
Eq.~(\ref{ddotdeltam}) gives \bea\label{deltama}
\delta_{m}^{\prime\prime}&+&\f{2ag^\prime+6g-3}{2ag}\delta_m^\prime-\left[\f{g+3}{3}\right]\f{\delta_m^{\prime2}}{1+\delta_m}\nn&-&\f{\kappa(1+\delta_m)(1-6\kappa\lambda)\rho_m\delta_m}{2ga^2H^2(1-4\kappa\lambda)}=0.
\eea
\section{Solutions in linear regime}\label{sollineareqs}
\begin{figure}[!]
    \includegraphics[width=8cm]{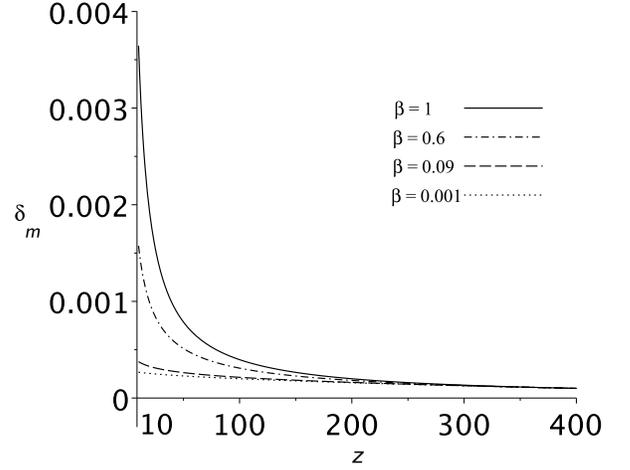}\vspace{0.4cm}
    \caption{Behavior of density contrast for different values of $\beta$ parameter during the evolution of the Universe in matter dominated era. We have set $\kappa_G=1$, $\alpha=3/7$ and $\delta_m(z_i)=0.001$.}\label{FIGWG1}
\end{figure}

\begin{figure}[!]
    \includegraphics[width=8cm]{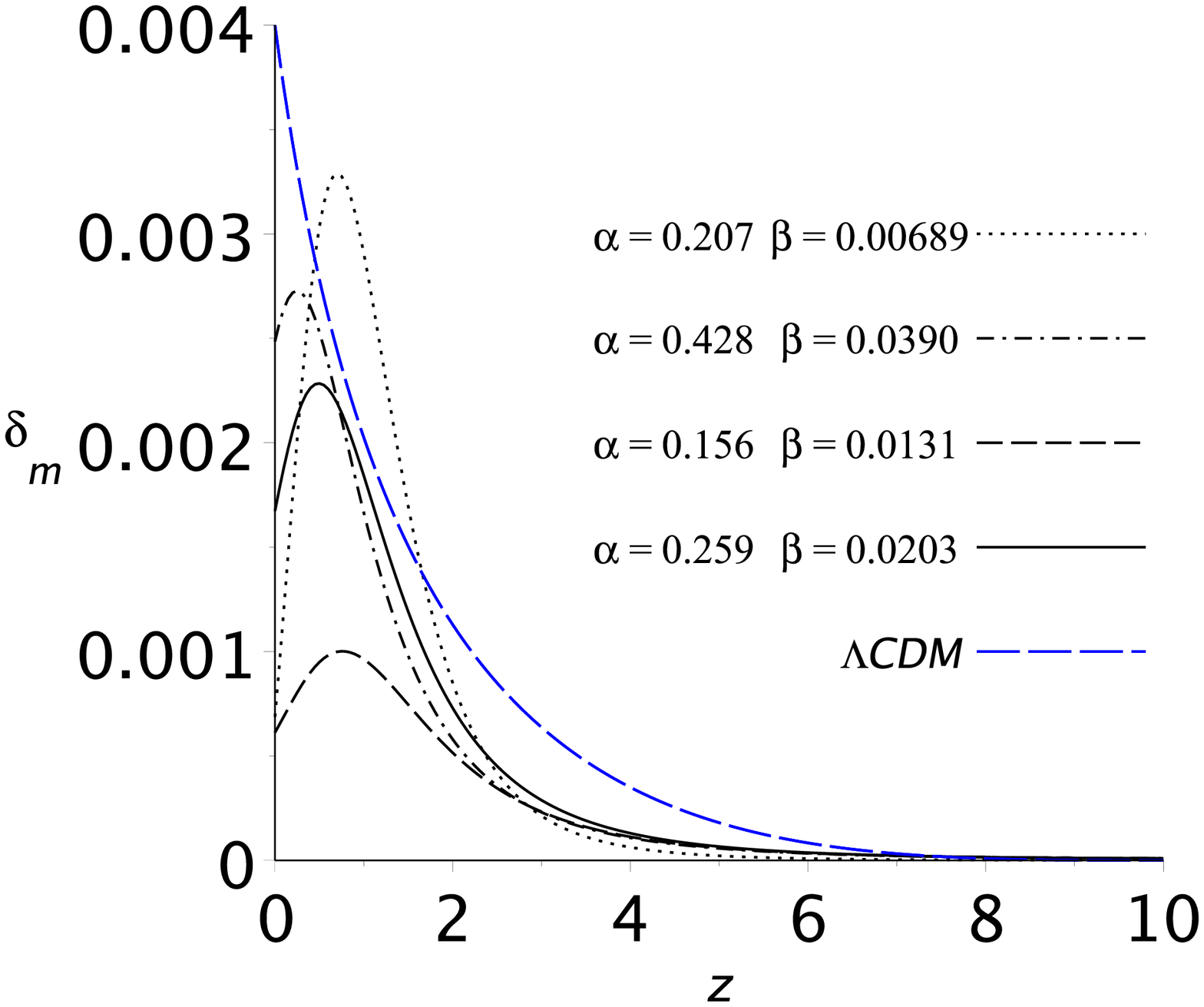}\vspace{0.4cm}
    \includegraphics[width=8cm]{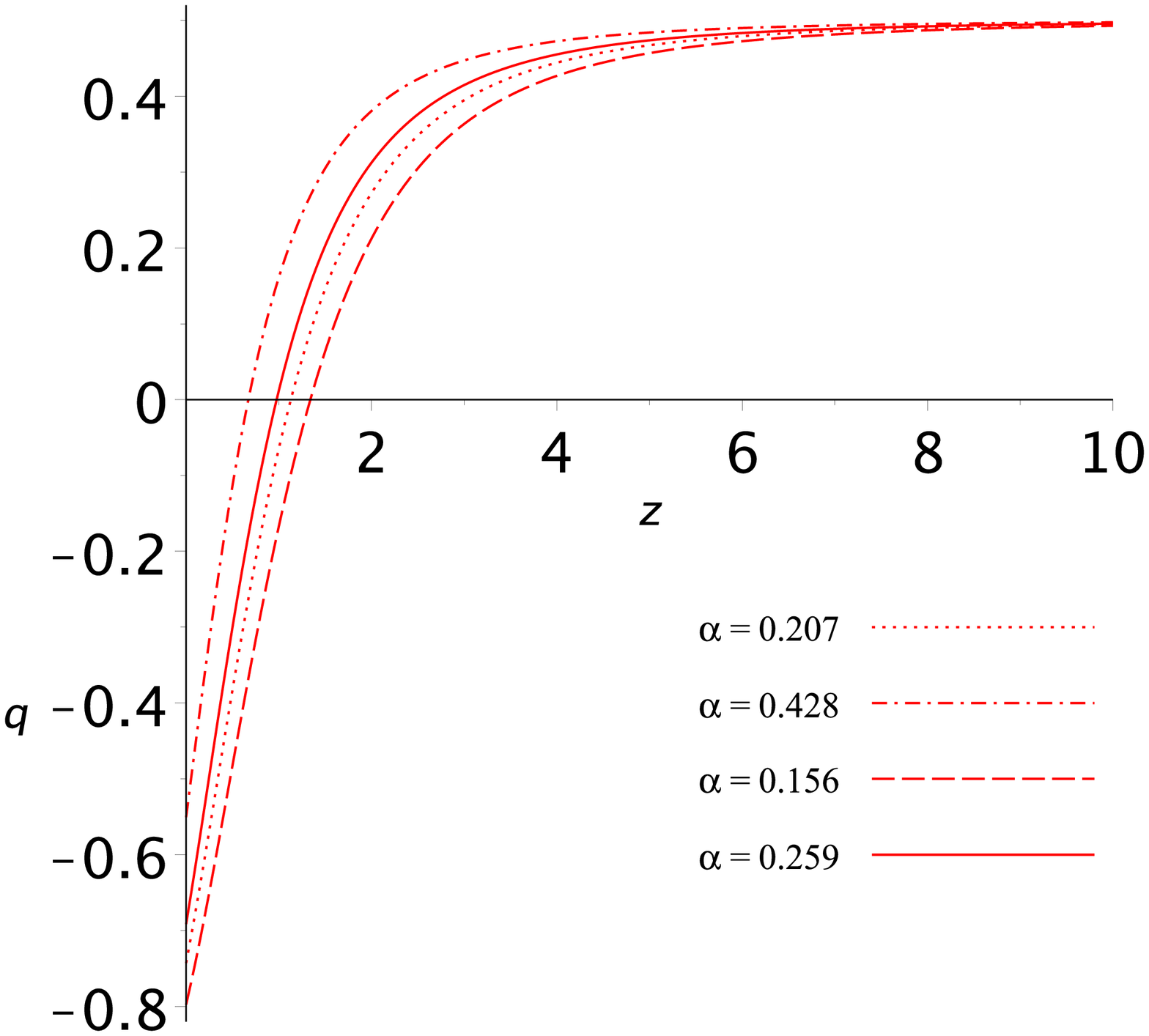}\vspace{0.4cm}
    \caption{Upper panel: Behavior of density contrast during the evolution of the Universe for different values of $\alpha$ and $\beta$ parameters. The maximums occur at the redshifts: $z_{\rm max}=0.707$ (dotted curve), $z_{\rm max}=0.263$ (dot-dashed curve), $z_{\rm max}=0.503$ (solid curve) and $z_{\rm max}=0.722$ (dashed curve). We have set $\kappa_G=1$, $\Omega_{m}^0=0.25$, $h_0=0.67$ and $\delta_m(z_i)=0.001$. Lower panel: Behavior of deceleration parameter for the same values of parameter $\alpha$ as the upper panel. The transition redshifts are given as: $z_{\rm tr}=1.129$ (dotted curve), $z_{\rm tr}=0.671$ (dot-dashed curve), $z_{\rm tr}=1.340$ (dashed curve) and $z_{\rm tr}=0.976$ (solid curve).}\label{FIGWG2}
\end{figure}

\begin{figure}[!]
    \includegraphics[width=8cm]{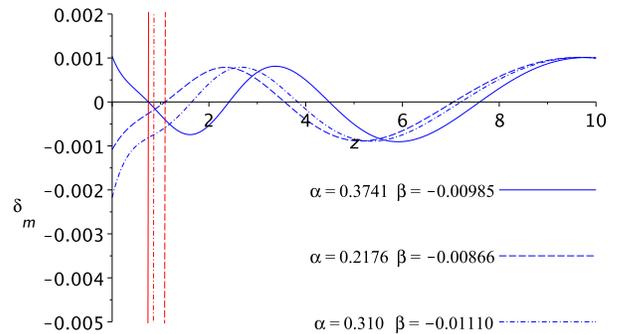}\vspace{0.9cm}
    \caption{Behavior of density contrast (blue curves) during the evolution of the Universe for different values of $\alpha$ parameter and negative values of $\beta$ parameter. The red curves denote the values of deceleration parameter between the interval $(-0.005,0.002)$. We have set $\kappa_G=1$, $\Omega_{m}^0=0.25$, $h_0=0.67$ and $\delta_m(z_i)=0.001$. The same value for parameter $\alpha$ has been considered for red curves.}\label{FIGWG3}
\end{figure}

In order to extract some physical results from Eq. (\ref{deltama}) we proceed with ignoring the terms containing ${\mathcal O}(\delta^2)$. We then get
\bea\label{deltamal}
\delta_{m}^{\prime\prime}+\left[\f{2ag^\prime+6g-3}{2ag}\right]\delta_{m}^\prime-\f{3\beta\delta_{m}\Omega_m(1-6\beta\lambda)}{2a^2g(1-4\beta\lambda)}=0,\nn
\eea
where we have set $\kappa=\beta\kappa_G$ with $\kappa_G=8\pi G$ being the Einstein's gravitational constant. The matter density parameter is given by
\be\label{mattdenp}
\Omega_m=\f{\kappa_G\rho_m}{3H^2}=\f{\alpha^2a^{-3}}{\beta^2\Omega_{m}^0H_0^2\left(1+\alpha a^{-3}\right)},
\ee
where use has been made of the solution for Hubble parameter~\ref{Hubblepar} and $\Omega_{m}^0$ and $H_0$ are the present values of density and Hubble parameters, respectively. Let us now  consider Eq.(\ref{deltamal}) for matter dominated epoch, i.e., $z\approx10^3$ where the matter density parameter can be approximated as $\Omega_{\rm m}\approx1$. By changing the independent variable from scale factor to redshift, equation (\ref{deltamal}) can be recast into the following form
\bea\label{deltamalz}
(1\!\!&+&\!\!z)^2\left(1+\alpha(1+z)^3\right)\f{d^2\delta_m(z)}{dz^2}\nn&+&\f{1}{2}(1+z)\left(-8+\alpha(1+z)^3\right)\f{d\delta_m(z)}{dz}\nn&-&\f{3\beta}{2}\left(-2+\alpha(1+z)^3\right)\delta_m(z)=0.
\eea
The above differential equation admits an exact solution given by
\bea\label{deltamalzsol}
\delta_m(z)&=&C_1(1+z)^{n_{-}}{}_2F_1\left[\ell_{+},\ell_{-},b_{+},-\f{1}{\alpha(1+z)^3}\right]\nn&+&C_2(1+z)^{n_{+}}{}_2F_1\left[s_{+},s_{-},b_{-},-\f{1}{\alpha(1+z)^3}\right],\nn
\eea
where
\bea\label{npbs+-}
n_{\pm}&=&\f{1}{4}\left(1\pm\sqrt{1+24\beta}\right),\nn
\ell_{\pm}&=&\f{3}{4}\pm\f{1}{6}\sqrt{25-12\beta}+\f{1}{12}\sqrt{1+24\beta},\nn
s_{\pm}&=&\f{3}{4}\pm\f{1}{6}\sqrt{25-12\beta}-\f{1}{12}\sqrt{1+24\beta},\nn
b_{\pm}&=&1\pm\f{1}{6}\sqrt{1+24\beta},
\eea
and $C_1$ and $C_2$ are integration constants. To obtain these two constants we note that for large values of the redshift the $\lambda$ parameter goes to zero where the GR limit is recovered. We therefore consider the adiabatic initial conditions for matter perturbation as
\be\label{adiainconds}
\f{d\delta_m(z)}{dz}\Bigg|_{z=z_i}\!\!\!=-\f{\delta_m(z_i)}{1+z_i},
\ee
where $\delta_m(z_i)$ is the initial value of density contrast at the onset of perturbations, i.e., at $z=z_i$. In GR limit where $\lambda\rightarrow0$ and $\beta\rightarrow1$ Eq. (\ref{deltamal}) can be re-expressed as
\be\label{GRLimit}
(1+z)^2\f{d^2\delta_m(z)}{dz^2}+\f{1}{2}(1+z)\f{d\delta_m(z)}{dz}-\f{3}{2}\delta_m(z)=0,
\ee
for which the solution reads
\be\label{GRLimitsol}
\delta_m(z)=\f{1+z_i}{1+z}\delta_m(z_i).
\ee
As it is expected, matter perturbations linearly grow with the scale factor~\cite{abramo2007}, however, these perturbations have different behavior in the framework of generalized Rastall gravity. In Fig. (\ref{FIGWG1}) we have plotted the evolution of matter density contrast for different values of $\beta$ parameter. The black solid curve represents matter density contrast for $\Lambda$CDM model, i.e., $\lambda=0$ and $\beta=1$~\cite{abramo2007}. Considering the solution (\ref{deltamalzsol}), the matter perturbations proceed with a smaller rate of growth, see the dotted, dashed and dot-dashed curves. Therefore, the inclusion of a dynamical coupling between matter and geometry suppresses the growth of matter perturbations
in this case. The solution (\ref{deltamalzsol}) may not be applied to $z<10$ till the present epoch. Considering then the dynamical behavior of density parameter (\ref{mattdenp}), Eq. (\ref{deltamal}) can be re-expressed as the following form
\bea\label{deltamalOV}
(1\!\!&+&\!\!z)\f{d^2\delta_m(z)}{dz^2}+\f{\left(-8+\alpha(1+z)^3\right)}{2\left(1+\alpha(1+z)^3\right)}\f{d\delta_m(z)}{dz}\nn&-&\f{3\alpha^2(1+z)^2\left(-2+\alpha(1+z)^3\right)\delta_m(z)}{2\left(1+\alpha(1+z)^3\right)^2H_0^2\Omega_m^0\beta}=0.
\eea
The above differential equation does not admit an exact analytical solution hence we resort to numerical methods. The upper panel in Fig. (\ref{FIGWG2}) shows the evolution of matter perturbations for redshifts $0\leq z\leq10$. The long-dashed blue curve presents the behavior of matter density contrast in $\Lambda${\rm CDM} model. It is therefore seen that matter perturbations for $\Lambda${\rm CDM} model grow monotonically as the Universe evolves to present times while, in the case of $\lambda\neq0$ these perturbations (black curves) start growing from their initial values and reach a maximum value at which the evolution of perturbations halts. This maximum occurs at a redshift, namely, $z_{\rm max}$. For $z<z_{\rm max}$ the density contrast decreases and reaches a constant nonzero value up until the present time. As the behavior of deceleration parameter (red curves within the lower panel) shows, the Universe experiences a transition from decelerating, $z<z_{\rm tr}$, to accelerating $z>z_{\rm tr}$ phases, where $z_{\rm tr}$ is the redshift at which the deceleration parameter vanishes. Comparing the two set of curves, we observe that $z_{\rm max}<z_{\rm tr}$. This means that as the Universe enters the accelerating phase, the rate of matter density contrast decreases since the accelerated expansion will produce a decrement in matter clustering. We note that the present time value of the matter density contrast is greater than its initial value i.e., $\lim_{\substack{z\rightarrow0}}\delta_m(z)>\delta_{m}(z_i)$. This means that, though the density contrast decays after transition redshift, we could have structure formation, even at the present evolution of the Universe. Another possible scenario is the oscillatory behavior of the density contrast which can be achieved by considering negative values of $\beta$ parameter. In this case, overdense and underdense regions within the Universe could form periodically so that, during the evolution of the Universe, formation of structures can occur at the maximum value of the density contrast and negative peaks imply the presence of voids in the matter distribution. Such a behavior occurs during a period of oscillation and at distinct redshifts $z_{\rm pr}$, at which $\delta_{m}(z_{\rm pr})=0$. The value of $z_{\rm pr}$ depends crucially on the absolute value of $\beta$ parameter. In Fig. (\ref{FIGWG3}) we considered three different scenarios (blue curves) for evolution of matter perturbations. The solid curve shows that after a cycle of formation of overdense and underdense regions, the matter perturbation starts growing from an underdense region toward the overdense ones. Such an event occurs at the transition redshift for which $q(z_{\rm tr})=0$, so that, $\delta_{m}(z)<0$ for $z>z_{\rm tr}$ and $\delta_{m}(z)>0$ for $z<z_{\rm tr}$. Hence we could have voids within the decelerated era and matter clustering in accelerating and present time eras. For the second scenario (dashed curve) we have overdense regions just before the transition redshift, i.e., $\delta_{m}(z)>0$ for $z>z_{\rm tr}$ and underdense regions after the transition is passed, i.e., $\delta_{m}(z)<0$ for $z<z_{\rm tr}$. Finally we could have underdense regions even before the transition redshift at which the Universe enters an accelerated regime (dot-dashed curve). In this case the perturbations turn into the formation of voids at the redshift $z_1>z_{\rm tr}$ for which $\delta_{m}(z_1)=0$. As the Universe evolves, the amplitude of perturbations grow in negative direction i.e., $\delta_{m}(z_{\rm tr})<0$ and reaches a finite negative value at present times.
\par
\begin{figure}[!]
	\includegraphics[width=8cm]{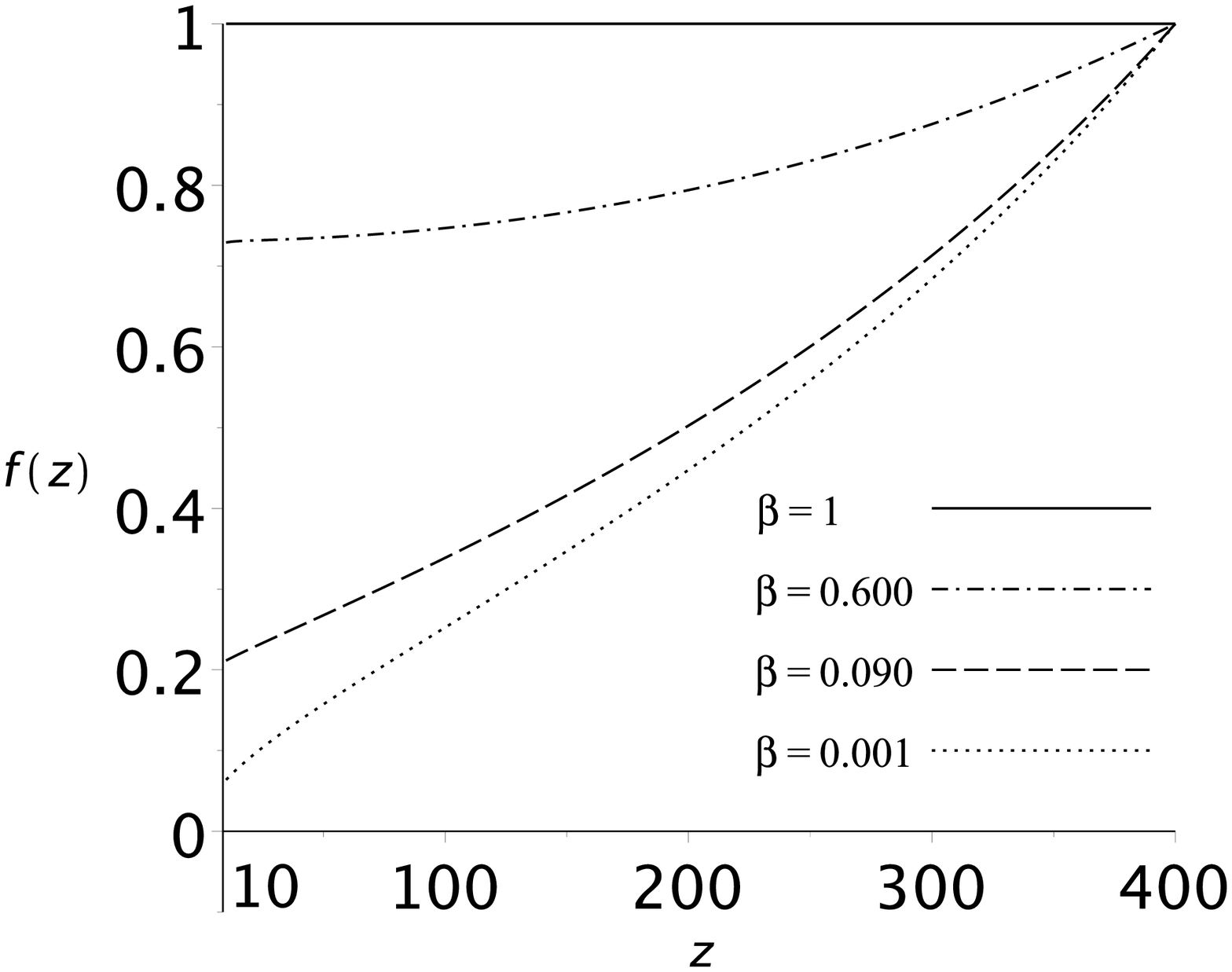}\vspace{0.4cm}
	\includegraphics[width=8cm]{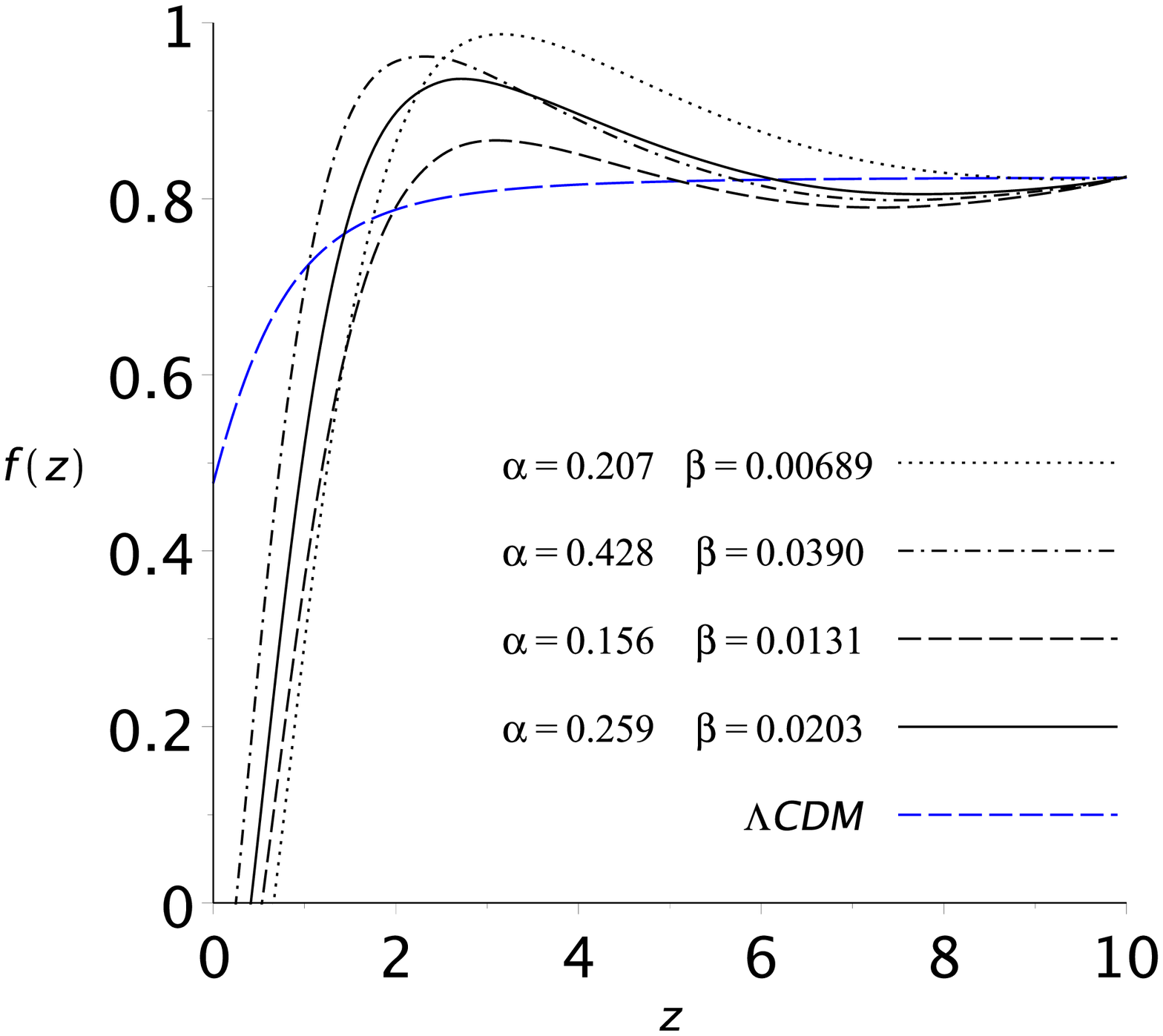}\vspace{0.42cm}
	\caption{Upper panel: Evolution of growth function during the matter dominated epoch for $\alpha=3/7$ and different values of $\beta$ parameter. We have set $\kappa_G=1$, and $\delta_m(z_i)=0.001$. Lower panel: Evolution of growth function for the density contrast as given in Fig. (\ref{FIGWG2}).}\label{FIG5}
\end{figure}

During the process of structure formation, the overdense regions grow with time due to the gravitational attraction. An estimation of the growth rate of matter perturbations is given by the growth function defined as~\cite{peebles1993}
\be\label{grfunction}
f(a)=\f{d\ln D}{d\ln a},~~~~D(a)=\f{\delta_m(a)}{\delta_m(a=1)}.
\ee
In matter dominated era, for $\lambda=0$ and $\beta=1$ the amplitude of perturbations obeys Eq. (\ref{GRLimitsol}) and thus the growth function is a constant of unity. However, for $\lambda\neq0$ the growth function behaves in a different manner. In the upper panel of Fig. (\ref{FIG5}) we have plotted for the behavior of growth function as a function of redshift. We observe that in the framework of generalized Rastall theory, the amplitude of perturbations grow but with a decreasing growth rate and the greater the value of $\beta$ parameter, the lesser the decrement in the growth function. In the lower panel, we have plotted the growth function for the case with varying matter density parameter. We therefore observe that, at higher redshifts, the growth function curve (family of black curves) for generalized Rastall model starts from the values close to the $\Lambda$CDM model (long-dashed blue curve). As the Universe evolves to lower redshifts, the growth function undergoes deviations from the $\Lambda$CDM curve. More precisely, the growth function increases for a while until reaching a maximum at $z=z^{\rm f}_{\rm max}$, after which it decreases to zero at the redshift $z=z_{\rm max}$, so that we have $f(z_{\rm max})=0$. Hence, regarding the upper panel of Fig. (\ref{FIGWG2}), we find that for $z>z^{\rm f}_{\rm max}$, the amplitude of perturbations grow in an accelerated way and for $z_{\rm max}<z<z^{\rm f}_{\rm max}$, the perturbations still continue to grow but in a decelerated way. 
\section{Non-linear Regime}\label{nonlinregime}
\begin{figure}[!]
    \includegraphics[width=8cm]{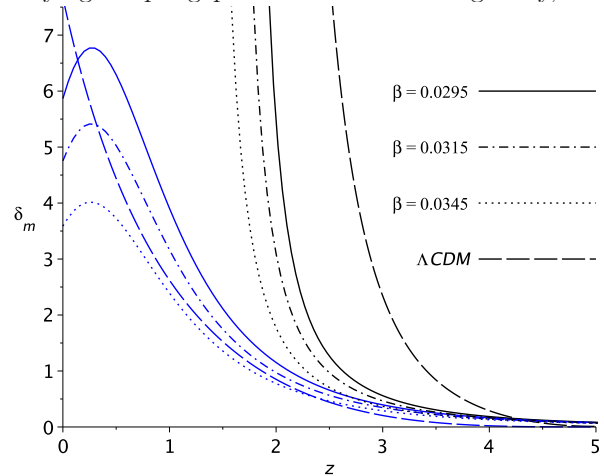}\vspace{0.4cm}
    \caption{Evolution of linear (blue curves, $\delta^{\rm L}_{m}$) and  non-linear (black curves, $\delta^{\rm NL}_{m}$) for $\alpha=3/7$ and different values of $\beta$ parameter. The long-dashed curves show the evolution of density perturbations of $\Lambda${\rm CDM} model in linear (blue) and non-linear (black) regimes. We have set $\kappa_G=1$, $\Omega_{m}^0=0.25$, $h_0=0.67$ and $\delta_m(z_i)=0.01$. The same value for parameter $\beta$ has been considered for the blue curves.}\label{FIGWG4}
\end{figure}

In this section we study the evolution of matter density contrast during the non-linear regime of structure formation. We therefore consider Eq. (\ref{deltama}) which in terms of redshift can be re-expressed as
\bea\label{eqdeltanonlin}
(1\!\!&+&\!\!z)^4\f{d^2\delta_{m}}{dz^2}+2(1+z)^3\f{d\delta_{m}}{dz}-\f{3}{2}(1+z)^3x(z)\f{d\delta_{m}}{dz}\nn&-&\f{3\alpha^2}{2\beta\Omega_{\rm m}^0H_0^2}(1+z^5)y(z)\delta_{m}(1+\delta_{m})\nn&-&\left[1+\f{1+\alpha(1+z)^3}{3\alpha(1+z)^3}\right]\f{(1+z)^4}{1+\delta_{m}}\left(\f{d\delta_{m}}{dz}\right)^2=0,
\eea
where
\bea\label{xyz}
x(z)&=&\f{\alpha z^3+3\alpha z^2+3\alpha z+\alpha+4}{\alpha z^3+3\alpha z^2+3\alpha z+\alpha+1},\nn
y(z)&=& \f{-2+\alpha(1+z)^3}{(1+\alpha(1+z)^3)^2}.
\eea
In Fig. (\ref{FIGWG4}) we have plotted numerical solution (black curves) of Eq. (\ref{eqdeltanonlin}) along with the results of linear regime (blue curves). It is seen that, perturbations grow as the Universe evolves and eventually the perturbations deviate from the linear regime at a critical redshift where the nonlinear regime starts to dominate the evolution of matter perturbations and gets arbitrary large values. We therefore observe that the collapse of matter occurs when a sufficient amount of density contrast is accumulated, i.e., when $\delta^{\rm NL}_{m}\rightarrow\infty$. From physical viewpoint, this scenario corresponds to the formation of an object such as a super-cluster. The larger the values of $\beta$ parameter the slower the growth of the matter perturbations and the lower the peak in density contrast of linear regime. Moreover, for larger values of $\beta$ parameter, the critical redshift decreases, hence, it takes longer time for perturbations to accumulate the critical amount of matter in order to trigger the collapse. In comparison to non-linear evolution of matter perturbations in $\Lambda${\rm CDM} model (long-dashed black curve), we observe that the effects of including a dynamical coupling between matter and geometry tends to delay the collapse scenario so that structures would have collapsed much earlier if the effects of a dynamical $\lambda$ parameter had not been taken into account.
\section{Concluding Remarks}\label{conclrems}
In the framework of recent generalization of Rastall gravity, it is found that a dynamic coupling parameter could act as an origin for DE, proving then a suitable setting for current accelerated phase of the Universe expansion. In the present work, motivated by this idea, we studied evolution of a pressure-less matter perturbations using spherical top-hat collapse model. The exact solutions in linear regime show that matter density contrast grows until reaching a maximum value at a certain redshift. It then decreases and reaches a finite positive value depending on model parameters. The redshift at which the amplitude of perturbations ceases to increase is smaller than the transition redshift. Such a decrement in the amplitude of perturbations could be due to the accelerated expanding phase the Universe is experiencing after transition redshift. For some choice of model parameters the perturbations show oscillatory behavior so that we could have a period of overdense and underdense regions at different redshifts during the evolution of perturbations. The matter perturbations grow faster in non-linear regime in comparison to the linear one and at a critical redshift, non-linear density contrast starts detaching from the linear one and diverges, thus, the collapse of a massive sphere occurs. We observed that in the standard $\Lambda${\rm CDM} model, such a collapse process takes place earlier than the case in which the effects of a varying $\lambda$ parameter are present, hence, a dynamical interaction between matter and geometry tends to delay the accumulation of matter and thus formation of large-scale structures. We therefore conclude that a running interaction between matter and geometry, represented by a varying coupling parameter in Rastall gravity, could affect the dynamics of matter perturbations and consequently formation of structures during the evolution of the Universe.
\section{Acknowledgment}
The authors would like to appreciate the anonymous referee for providing useful comments and suggestions that helped us to improve the original version of our manuscript.
\end{document}